\documentclass[12pt, letterpaper]{article}

\usepackage{amsmath, amsfonts}
\usepackage{booktabs}
\usepackage{amsthm}
\usepackage{graphicx}
\usepackage[margin=1in]{geometry}
\usepackage{hyperref}
\usepackage{cleveref}
\hypersetup{colorlinks = true, linkcolor = blue, citecolor=blue, urlcolor = blue}
\usepackage{natbib}
\usepackage{float}
\usepackage{setspace}
\usepackage{pdfpages}
\usepackage{lineno}
\usepackage{mwe}
\usepackage{comment}

\newcommand*\patchAmsMathEnvironmentForLineno[1]{%
 \expandafter\let\csname old#1\expandafter\endcsname\csname #1\endcsname
 \expandafter\let\csname oldend#1\expandafter\endcsname\csname end#1\endcsname
 \renewenvironment{#1}%
 {\linenomath\csname old#1\endcsname}%
 {\csname oldend#1\endcsname\endlinenomath}}%
\newcommand*\patchBothAmsMathEnvironmentsForLineno[1]{%
 \patchAmsMathEnvironmentForLineno{#1}%
 \patchAmsMathEnvironmentForLineno{#1*}}%

\AtBeginDocument{%
 \patchBothAmsMathEnvironmentsForLineno{equation}%
 \patchBothAmsMathEnvironmentsForLineno{align}%
 \patchBothAmsMathEnvironmentsForLineno{flalign}%
 \patchBothAmsMathEnvironmentsForLineno{alignat}%
 \patchBothAmsMathEnvironmentsForLineno{gather}%
 \patchBothAmsMathEnvironmentsForLineno{multline}%
}

\newcommand{\eds}[1]{\textcolor{red}{EDS: (#1)}}
\newcommand{\of}[1]{\textcolor{violet}{OF: #1}}

\newcommand{\blind}{0}

\begin{document}

\title{\bf On Devon Allen's Disqualification at the 2022 World Track and Field Championships}

\if0\blind
{
  \author{Owen Fiore, 
  Elizabeth D. Schifano, 
  Jun Yan\\[1ex]
  Department of Statistics, University of Connecticut\\
}
} \fi

\if1\blind
{
  \bigskip
  \bigskip
  \bigskip
  \author{Anonymous Authors}
  \bigskip
} \fi

\maketitle

\doublespace


\begin{abstract}
Devon Allen’s disqualification at the men's 110-meter hurdle final at
the 2022 World Track and Field Championships, 
due to a reaction time (RT) of 0.099 seconds---just 0.001 seconds below
the allowable threshold---sparked widespread debate over 
the fairness and validity of RT rules. This study investigates two key
issues: variations in timing systems and the justification for the
0.1-second disqualification threshold. We
pooled RT data from men’s 110-meter hurdles and 100-meter dash, as
well as women’s 100-meter hurdles and 100-meter dash, spanning
national and international competitions. Using a rank-sum test for
clustered data, we compared RTs across multiple competitions,
while a generalized Gamma model with random effects for venue and heat
was applied to evaluate the threshold. Our analyses reveal significant
differences in RTs between the 2022 World Championships and other
competitions, pointing to systematic variations in timing
systems. Additionally, the model shows that RTs below 0.1 seconds,
though rare, are physiologically plausible. These findings highlight
the need for standardized timing protocols and a re-evaluation of the
0.1-second disqualification threshold to promote fairness in
elite competition.

\end{abstract}

\doublespace

\section{Introduction}
\label{sec:intro}

Devon Allen’s highly anticipated performance at the 2022 World
Track and Field Championships in Eugene, Oregon, ended in
controversy when he was disqualified for a reaction time (RT) of
0.099 seconds, just 0.001 seconds below the allowable threshold.
Allen, a University of Oregon alumnus,
had recently run a time of 12.84 seconds in the 110-meter hurdle
event, just 0.04 seconds short of the world record. After placing
third at the U.S. Track and Field Championships, he advanced
through the preliminary heats and semifinals at the World
Championships, with RTs of 0.123 and 0.101 seconds,
respectively. However, in the final heat, competing in front of his
home audience, Allen’s RT was just 0.001 seconds faster
than the 0.1-second threshold set by the International Association
of Athletics Federations (IAAF).  His resulting disqualification was
met with widespread public outcry. This 
incident highlighted two long-standing issues: variability in the
measurement of RTs by Start Information Systems (SIS) and the
appropriateness of the 0.1-second disqualification threshold. As
RTs are measured in fractions of a second, inconsistencies
in timing technologies and rules can significantly affect athlete
outcomes, raising questions about fairness and standardization.

World Athletics (formerly IAAF) uses certified SIS to measure RTs, yet
variation in technology persists. Discussions at online forums such as
\url{www.LetsRun.com} questioned consistencies in the SIS as a
contributing factor to RT anormalies \citep{johnson2022data,
  johnson2022was}. Historically, ``loud gun'' systems caused signal delays
for athletes in outer lanes  due to the speed of sound, an issue
addressed with the introduction of ``silent gun'' systems in 2010,
which electronically synchronize sound delivery to all athletes
\citep{tonnessen2013reaction}. Despite these advances, variability
persists due to differences in sensor technologies, such as force
transducers and accelerometers, and inconsistencies in event detection
algorithms \citep{willwacher2013novel}. For example, simple
force-threshold systems may delay RT detection by up to 26~ms compared
to more sophisticated methods \citep{pain2007sprint}. These findings
emphasize the need for standardized certification protocols to reduce
discrepancies and ensure fairness in RT measurements, as recently
reviewed by \citet{milloz2021sprint}.

Originally introduced in the 1990s to discourage athletes from attempting
to predict the start gun, the 0.1-second disqualification threshold 
has been the subject of significant debate.   Partly based on limited
data from Finnish national-level athletes~\citep{mero1990reaction}, the
threshold may not adequately represent the capabilities of elite
sprinters. Controlled experiments have shown that RTs below
0.1 seconds are physiologically plausible \citep{pain2007sprint,
  komi2009iaaf}, while retrospective analyses of competition data
often advocate for raising the threshold~\citep{brosnan2017effects,
  lipps2011implications}. Stricter false-start rules, introduced to
minimize race disruptions, have also discouraged sprinters from
attempting faster starts, which may artificially inflate RTs recorded
in competition~\citep{haugen2013effect}. This divergence between
experimental findings and competition-based analyses illustrates the
complexity of defining a universally fair threshold. Addressing these
debates requires modern data collection and advanced methodologies to
ensure equity and consistency in elite
competition~\citep{milloz2021sprint}.

This paper addresses two primary objectives from a statistical
perspective, using modern methodologies to analyze historical
data. First, we investigate whether RTs at the 2022
World Championships were significantly different from other
competitions, focusing on athletes who competed in multiple
events. Using a matched-pairs design, we compare RTs across the 2022,
2019, and 2023 World Championships, as well as 2022 national-level
competitions. This approach isolates the effect of the competition
 while controlling for individual performance. With the goal of assessing 
differences across competitions within athletes, RTs were analyzed using a 
rank-based comparison approach for clustered data \citep{datta2005rank}. Second, 
we evaluate
the appropriateness of the 0.1-second disqualification threshold by
modeling RTs from World Championships held from 1999 onward. A
generalized Gamma (GG) distribution with random effects for both venue and
heat was applied within the framework of the generalized additive
model for location, scale, and shape (GAMLSS)
\citep{rigby2005generalized, stasinopoulos2024generalized}.
This model enables estimation of the
probability of RTs falling below a threshold, providing a
statistical assessment of RT consistency and the validity of the
current threshold.

The rest of this paper is organized as follows. Section~\ref{sec:problem1}
investigates RTs of athletes who competed at both the 2022 World Championships 
and another competition to examine differences between their RTs.  
Section~\ref{sec:problem2}
investigates RTs of athletes from 1999 to 2023 to determine a reaction barrier
ground in statistical analysis.  Within each of the above sections, the data,
methods used, and results are presented.
Finally, Section~\ref{sec:disc} highlights the
paper’s impact, limitations, and potential for future research.
All data and code for our analysis are provided in the Supplementary
Materials.

\section{Assessing the 2022 World Championships RTs}
\label{sec:problem1}

The notably faster RTs at the 2022 World Championships
raised concerns about systematic biases. 
For example,  the median RTs across multiple sprint events were found
to be the lowest in recent history; the number of RTs recorded under
0.115 at men’s 100m dash and 110m hurdles was 25, much greater than that in
2019, which was 3 \citep{johnson2022was}. This striking disparity
suggests a potential systematic difference in RT measurements at the
2022 World Championships, necessitating a formal statistical
investigation. We assess whether RTs at the 2022 World Championships
were systematically faster by comparing them against (1) RTs from national
competitions in 2022, (2) RTs from the 2019 World Championships, and (3) RTs 
from the 2023 World Championships, using data from athletes who competed in both 
competitions in the pairwise comparison.

\subsection{Data}
\label{sec:data_2022}

To investigate whether RTs at the 2022 World Championships
were significantly different from other competitions, we
used data from male athletes in the 110-meter hurdles and 100-meter
dash, and female athletes in the 100-meter hurdles and 100-meter dash,
provided they competed in the 2022 World Championships and at least
one other competition (2022 national championships, 2019 World
Championships, or 2023 World Championships). Statistical tests showed no
significant differences in RTs between these short hurdel and dash events, 
supporting their inclusion in a unified analysis. However, we excluded data from
200-meter dashes and longer events, as their RTs were found
to be significantly different. This exclusion is expected, as RT plays a smaller 
role in longer sprint events, where acceleration off the blocks is less decisive.
Negative RTs were excluded from the analysis, while positive disqualified RTs
were included due to their low frequency and not being obvious outliers.

\subsubsection{2022 National Competitions}
\label{sec:datanational}

We first compare RTs from the 2022 World Championships to national
competitions earlier that year to assess whether the same athletes
reacted differently across competitions.
Prior to a formal analysis, we examined how United States (US) athletes
performed at the 2022 US Track and Field Championships, held from June 23--26,
2022, at Hayward Field in Eugene, Oregon. Since this venue also hosted
the 2022 World Championships in August, it
provided a unique opportunity to compare. Among the four US
110-meter hurdle athletes---Trey Cunningham, Daniel Roberts,
Grant Holloway, and Devon Allen---all
recorded faster RTs in every World Championships race
compared to their performances at the national-level event. Similarly,
all four US 100-meter dash athletes---Marvin Bracy, Fred Kerley,
Travyon Bromwell, and Christian Coleman---also recorded faster RT's
at the World Championships, reinforcing concerns 
about systematic differences in reaction measurements.

To expand the dataset, we included RTs from 100-meter hurdle
and 100-meter dash athletes who competed in other national
competitions held between May and July 2022 across various
countries. Compiling this data presented challenges, as RTs were not
centrally archived and often required searching country-specific
websites, with many results recorded in native languages.

\begin{figure}[tbp]
  \centering
  \includegraphics[width=\textwidth]{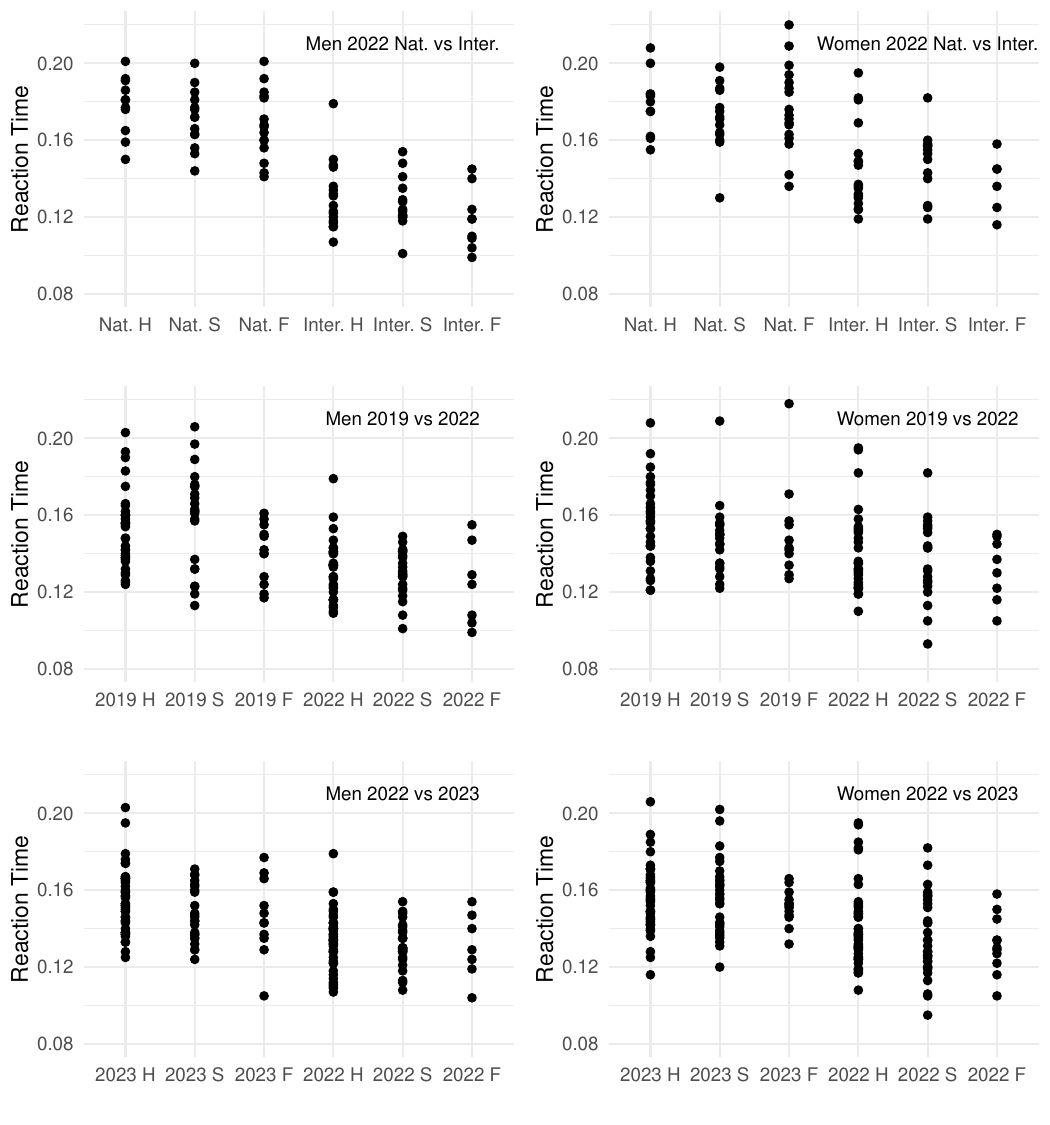}
  \caption{RTs for athletes who competed at the 2022 World Championships
    and at another championship (2022 national, 2019 World, or 2023
    World) at which they competed. On the horizontal
    axis below each graph ``H'', ``S'', and ``F'' refer to the heats,
    semifinals, and  finals respectively. Please note that in the last
    row the 2022 times are to the right of the 2023 times.}
  \label{fig:RankScatterplots}
\end{figure}

The final dataset consisted of RTs from athletes who competed
in both a 2022 national competition and the international 2022 World Championships.
RTs from preliminary heats, semifinals, and finals were included to ensure
sufficient data for analysis. Excluding preliminary heats would have
significantly reduced the number of athletes and clusters. Each
athlete was considered as a cluster, with observed RTs
from both the `treatment' group (2022 World Championships) and
the `control' group (national competition) within the cluster. Cluster
sizes ranged from three to six, with a median size of four. Because gender
is known to influence RTs \citep{babicc2009reaction,
  lipps2011implications}, we prepared data for men and women
separately, resulting in 80 RTs from 17 athletes for each
gender. This structure enables a rank-based comparison for clustered 
data, which properly accounts for the within-athlete dependence inherent 
in this dataset. The top
panel of Figure~\ref{fig:RankScatterplots} shows the RTs
of those who competed both at national competitions and the 2022 World
Championships.

\subsubsection{2019 and 2023 World Championships}
\label{sec:data2019}

To determine whether RTs at the 2022 World Championships were
anomalous within the World Championships series, we compare them to
RTs from the 2019 and 2023 events. In the 2019-2022 comparison,
RTs from 2022 were considered as the `treatment' group, with
2019 serving as the `control' group. Similarly, in the 2022-2023
comparison, RTs from 2022 were considered as the `treatment'
group, with 2023 serving as the `control' group. This structure allowed
us to prepare datasets suitable for examining RTs of athletes
who competed across multiple World Championships.

Each athlete was treated as a single cluster, containing their RTs
from different World Championships. The dataset for the 2019-2022
comparison contained 134 RTs from 34 male athletes and 124
RTs from 31 female athletes. The dataset for the 2022-2023
comparison contained 161 RTs from 45 male athletes and 182
RTs from 47 female athletes. While it is theoretically
possible that athletes improved their RTs between 2019 and
2022 or between 2022 and 2023, such improvements are highly unlikely for
elite sprinters, as they already operate near the limits of human
performance. Consequently, consistent improvements observed in 2022
would suggest systematic differences rather than natural variability.

Figure~\ref{fig:RankScatterplots} shows the RTs of athletes
who competed in both the 2019 and 2022 World Championships (middle
panel) and those who competed in both the 2022 and 2023 World
Championships (lower panel). Since the sets of athletes differ between 
the two comparisons, each analysis provides independent evidence 
for evaluating potential anomalies in 2022 RTs. Notably, Devon Allen
recorded the fastest RTs in both the Finals and Semifinals of the
2022 World Championships, but his disqualification was determined by a
difference of just 0.002 seconds, with RTs of 0.101 and
0.099 seconds, respectively. This highlights the critical role of
RT precision in elite-level competition.

\subsection{Methods}
\label{sec:methods_2022}

To test whether the timing system at the 2022 World Championships 
systematically produced faster RTs, we compare RTs recorded at this
event against those from the same athletes in other competitions. In
this setting, we have clustered data with subunit grouping. As discussed in the
previous section, each athlete serves as a cluster, with multiple RTs recorded 
from the same athlete across different competitions.
Let $X_{ij}$ be the $j$th RT of athlete~$i$, $i = 1, \ldots, n$,
$j = 1, \ldots, m_i$ where $m_i$ is the number of observations from
athlete~$i$. Let $\delta_{ij}$ be the group indicator of $X_{ij}$; $\delta_{ij}
= 1$ if $X_{ij}$ is in group~1 (2022 World Championships) and $\delta_{ij} = 0$
otherwise. Athletes are assumed to be independent, while subunit observations 
from the same athlete are not. The null hypothesis $H_0$ to be tested is that there is no difference
between the two groups; i.e., the distribution of $X_{ij}$ remains the same
regardless of the group indicator $\delta_{ij}$.

\citet{datta2005rank} proposed an extension of the Wilcoxon rank-sum test to
clustered data with subunit-level grouping. The test is based on a
within-cluster resampling approach that preserves the within-cluster
dependence. Consider randomly picking one observation
from each cluster to form a pseudo-sample. Let $X_i^*$ be a random pick from the
$i$th cluster in the pseudo-sample and $\delta_i^*$ its group indicator. The
Wilcoxon rank-sum statistic for the pseudo-sample is
\[
W^* = \frac{1}{n + 1} + \sum_{i=1}^{n} \delta_{i}^{*} R_{i}^{*},
\]
where $R_{i}^{*}$ is the rank of $X_{i}^{*}$ in the pseudo-sample.
The test statistic $S$ is the average of $W^*$, averaged over all possible
pseudo-samples conditioning on the observed data and group indicators.
The mean and variance of $S$ under $H_0$ can be derived so that $S$ can be
standardized to form a $Z$ statistic which follows a standard normal distribution
asymptotically \citep[p.910]{datta2005rank}.

For small sample sizes, the asymptotic normality may be unreliable. To
address this, we also use 1~million random permutations to
simulate the null distribution of the test statistic.
This method is available from the \texttt{clusWilcox.test()} function
with \texttt{method = `ds'} (for \underline{D}atta and \underline{S}atten) and
\texttt{exact = TRUE} from R package
\texttt{clusrank}; with \texttt{exact = FALSE}, the same function implements the 
asymptotic rank-based test \citep{jiang2020wilcoxon}.

\subsection{Results}
\label{sec:results_2022}

The rank-based methods described in Section~\ref{sec:methods_2022} were used to
compare RTs between the 2022 World Championships and other
competitions in which the same athletes participated. These comparisons
were conducted separately for men and women, resulting in six total
comparisons: RTs from the 2022 national-level championships
versus the 2022 World Championships for men and women, RTs
from the 2019 versus 2022 World Championships for men and women, and
RTs from the 2022 versus 2023 World Championships for men and
women.

\begin{table}
  \centering
  \caption{P-values of comparisons between
    RTs from different competitions for the same athletes.
    2022 Nat. vs Inter. compares RTs from 2022 national-level
    championships and the 2022 World Track and Field Championships. 2019
    vs 2022 compares RTs from the 2019 and 2022 World Track and
    Field Championships. 2022 vs 2023 compares RTs from the
    2022 and 2023 World Track and Field Championships.}
  \begin{tabular}{c c c c c}
   \toprule
   Comparison & Permutation & Asymptotic & \# of athletes & \# RTs  \\
   \midrule
   2022 Nat. vs Inter. Men & $1.0 \cdot 10^{-6}$ & $ 6.1 \cdot 10^{-5}$ & 17 & 80 \\
   2022 Nat. vs Inter. Women & $1.0 \cdot 10^{-6}$ & $ 1.2 \cdot 10^{-3}$ & 17 & 80 \\[1ex]
   2019 vs 2022 Men & $2.8 \cdot 10^{-5}$ & $1.1 \cdot 10^{-5}$ & 34 & 134 \\
   2019 vs 2022 Women & $ 1.5 \cdot 10^{-3}$ & $6.9 \cdot 10^{-3}$ & 31 & 124 \\[1ex]
   2022 vs 2023 Men & $1.0 \cdot 10^{-6}$ & $1.4 \cdot 10^{-6}$ & 45 & 161 \\
   2022 vs 2023 Women & $1.0 \cdot 10^{-6}$ & $9.4 \cdot 10^{-7}$ & 47 & 182 \\
   \bottomrule
  \end{tabular}
  \label{tab:Clusrankresults}
\end{table}

Table~\ref{tab:Clusrankresults} presents the results from both permutation
and asymptotic rank-based tests for the six comparisons. All tests
yielded very small p-values, even after applying a Bonferroni
adjustment for multiple comparisons, indicating consistent evidence of
faster RTs at the 2022 World Championships relative to other
competitions. For both men and women, the national versus
international comparisons showed that RTs at the 2022 World
Championships were significantly faster than at national-level
competitions held earlier that year. Similarly, comparisons between
the 2019 and 2022 World Championships and between the 2022 and 2023
World Championships produced significant results, reinforcing the
observation that RTs at the 2022 World Championships were notably
faster. These findings support the hypothesis that conditions at the
2022 World Championships, whether systematic or
environmental, contributed to consistently faster RTs.

We also conducted the same analysis with men’s and women’s data
pooled, yielding similar results. Details are provided in Section~1 of
the Supplementary Material.

\section{Evaluating The 0.1 Second RT Threshold}
\label{sec:problem2}

This section evaluates whether the 0.1-second RT threshold remains a valid 
disqualification criterion in elite sprinting. Using historical RT data from 
World Championships since 1999, we fit a generalized Gamma model with 
venue- and heat-level random effects to estimate the probability of RTs below 
0.1 seconds. This statistical framework allows us to assess whether an 
alternative threshold would better align with observed RT distributions.

\subsection{Data}
\label{sec:data_barrier}

The data for evaluating the appropriateness of the 0.1-second
threshold was obtained from World Athletics and covers the men's
110-meter hurdles and 100-meter dashes from 1999 to 2023. Due to
possible gender differences \citep{babicc2009reaction,
  lipps2011implications}, data for women's 100-meter hurdles and
100-meter dashes were collected over the same time period with the analogous 
analyses relegated to the Supplementary Material. We focus on the RTs recorded during
semifinal and final heats only, as RTs from preliminary
heats are often not as fast as those in later heats
\citep[e.g.,][]{collet1999strategic, tonnessen2013reaction,
  brosnan2017effects, zhang2021correlation}. For analysis purposes, we
pooled RTs from semifinal and final heats to increase sample
size, which is particularly important for years with limited final heat
observations. For example, in 2022, only five data points were available
from the final heat due to two disqualifications and one athlete not
competing. Unless otherwise noted, this pooled dataset forms the basis
for our analyses for Objective 2. Additionally, we consider datasets
that exclude 2022 to assess how our findings might differ when excluding
this year of interest. This investigation began shortly after the 2022 World
Championships, and we were pleased that including data from 2023 did not
significantly alter our results \citep{WAData}.

\begin{figure}[tbp]
  \centering
  \includegraphics[width=\textwidth]{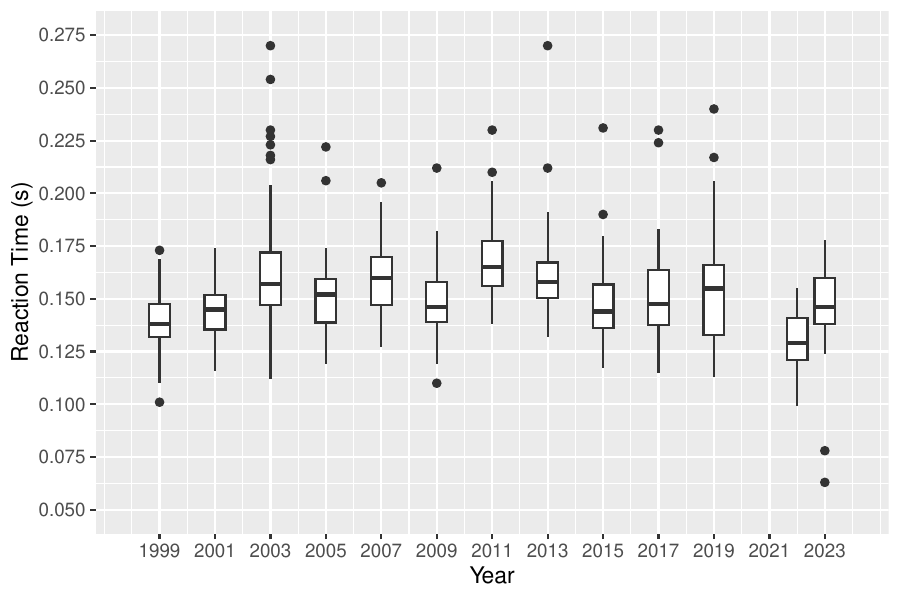}
  \caption{The RTs from 1999 to 2023 for the men's 110 meter hurdle
  and 100 meter dash.}
  \label{fig:Boxplot}
\end{figure}

The data is summarized in Figure~\ref{fig:Boxplot}, which presents a
sequence of boxplots of RTs from 1999 to 2023. It is evident
that RTs in 2022 were notably faster, with a median RT of 0.129 seconds compared
to the 0.156 seconds observed in earlier studies, such as
\citet{brosnan2017effects} for data spanning 1999 to 2014.
Figure~\ref{fig:Boxplot} also highlights year-to-year variability in
RTs, likely influenced by changes in the championship venue
and environmental conditions such as humidity, precipitation, and
elevation. Furthermore, advancements in technology and alterations to
false start rules during the study period may have played a role in these
variations \citep{willwacher2013novel}.

Between 2007 and 2009, World Athletics allowed one
false start warning before disqualifying a sprinter \citep{iaaf2009falsestart}.
This lenient rule led to 18 male and 7 female false starts at both the 2007
and 2009 World Championships. In 2011, this rule was replaced with the stricter
policy of automatic disqualification for false starts, aimed at reducing the
delays caused by repeated warnings. This change reduced men’s false starts
by two-thirds in 2011, with only six male and four female disqualifications
\citep{iaaf2009falsestart}. \citet{haugen2013effect} demonstrated that more
lenient false start rules significantly improved RTs during the
1997–2009 period, suggesting that rule changes over the study period may
have contributed to variations in RTs across years.

\subsection{Methods}
\label{sec:methods_barrier}

Based on an exploratory analysis, the RTs are adequately
modeled by a GG distribution with random effects in
model parameters. The GG distribution has three parameters, denoted by
$\text{GG}(\mu, \sigma, \nu),$ and has density function
\begin{equation}
  \label{eq:gg}
f_Y(y \mid \mu, \sigma, \nu) =
\frac{|\nu| \theta^\theta z^{\theta}}{\Gamma(\theta) y}
\exp\left(-z \theta\right),
\end{equation}
for $y > 0$, $\mu > 0$, $\sigma > 0$, and $\nu \neq 0$,
where $z = (y / \mu)^\nu$,
$\theta = 1 / (\sigma^2 \nu^2)$, and
$\Gamma(\cdot)$ denotes the Gamma function.
The GG distribution is highly flexible, encompassing several
well-known distributions as special cases, such as the
Weibull ($\mu = \nu)$ and  Gamma $(\nu = 1)$ distributions.
Its expectation is
\[
  \frac{\mu \Gamma(\theta + 1 / \nu)}
  {\theta^{1 / \nu} \Gamma(\theta)},
\]
provided $\theta > -1 / \nu$. Here,
$\mu$ scales the central tendency, $\sigma$ controls
dispersion, and $\nu$ determines skewness. This parameterization allows
the distribution to model asymmetric and heavy-tailed data effectively, making
it particularly suitable for RTs.
An implementation of this distribution is available from R package
\texttt{gamlss.dist} \citep{rigby2019distributions}.

Random effects at the venue and heat levels are incorporated into the
parameters of the GG distribution in~\eqref{eq:gg}.
Let $Y_{ijk}$ denote the RT of observation~$k$ in heat~$j$
of year~$i$. Conditioning on a venue effect $v_i$ for year~$i$
and a heat effect $h_{i/j}$ nested within each year~$i$, the
distribution of $Y_{ijk}$ is
$\text{GG}(\mu_{ijk}, \sigma_{ijk}, \nu)$, where
\begin{align}
\log(\mu_{ijk}) &= \beta_0 + v_i , \label{eq:mu}\\
\log(\sigma_{ijk}) &= \gamma_0 + h_{i/j} , \label{eq:sigma}
\end{align}
$v_i$ is normally distributed with mean zero and
variance~$\tau_v^2$, and $h_{i/j}$ is normally distributed with mean
zero and variance~$\tau_h^2$.
The two random effects were found useful: one 
capturing the venue effect, which is used to contrast years, and the 
second being the heat effect, where every race was given a unique 
identifier with typically five to nine observations per race.
This model can be fit with R package \texttt{gamlss} 
\citep{stasinopoulos2008generalized}. The heat effect could be added
to the model for $\mu_{ijk}$ and the venue effect could be added to 
the model for $\sigma_{ijk}$. From our comparison using the Akaike 
Information Criterion, Models~\eqref{eq:mu}--\eqref{eq:sigma} turned 
out to be preferred to more complex models or competing models.

Model diagnosis and tail analysis can be done with the fitted GG model
from package \texttt{gamlss}. Normalized quantile residuals, or
z-scores \citep{dunn1996randomized}, of the observations can be
extracted with the \texttt{residuals} method of a \texttt{gamlss}
object. The z-scores can then be checked with a Q-Q plot
\citep{almeida2018ggplot2}. The marginal
distribution of $Y_{ijk}$ is a scale-mixture of GG distributions, which can be
easily simulated from once the parameters are estimated. Many
random numbers generated from the fitted mixture distribution can be used to
approximate the probability of observing a RT faster than any given
threshold. We are specifically interested in the probability of a RT
being less than 0.1 seconds in order to gauge if that is a reasonable
disqualification barrier.

\subsection{Results}
\label{sec:results_barrier}

\begin{table}[b]
  \centering
  \caption{Estimated fixed-effect parameters with standard errors in
    parentheses and estimated standard deviations of the random effects from the
    fitted GG distribution with venue level random
    effects in $\mu$ and heat level random effects in $\sigma$ in
    Models~\eqref{eq:gg}--\eqref{eq:sigma}.}
  \label{tab:ggfit}
  \begin{tabular}{c c c c c c}
    \toprule
    Data set & $\beta_0$ & $\gamma_0$ & $\nu$ & $\tau_v$ & $\tau_h$ \\
    \midrule
    Excluding 2022 & $-$1.910 (0.005) & $-$2.200 (0.025) & $-$1.177 (0.442) & 0.043 & 0.326 \\
    Including 2022 & $-$1.910 (0.005) & $-$2.200 (0.027) & $-$1.178 (0.447) & 0.058 & 0.320 \\
    \bottomrule
  \end{tabular}
\end{table}

The results reported in this subsection are from men's data only
because our investigation found a significant gender difference.
Results for women's data are reported in Section~2 of the
Supplementary Material. The fitted parameters of the GG distribution in the
GAMLSS framework in Equations~\eqref{eq:gg}--\eqref{eq:sigma} are
summarized in Table~\ref{tab:ggfit}. Results obtained from both
excluding and including 2022 data are reported. The fixed-effect
parameters include $\beta_0$, $\gamma_0$, and $\nu$, corresponding to
the intercept of the log-location, log-scale, and shape of the GG
distribution, respectively. Random effects account for variability at
the venue level on the log-scale of the $\mu$ parameter
and at the heat level on the log-scale of the $\sigma$
parameter in the density in Equation~\eqref{eq:gg}. The variance
of the venue-level random effect is smaller than the heat-level random
effect variance, suggesting that heat-level variability in the scale
parameter is substantial, though on the dispersion parameter. When the
2022 data is included, all parameter estimates remain stable except
the standard deviation of the venue-level random-effect,
which increases from 0.043 to 0.058. These results
highlight that RTs are influenced by both venue and heat-level
factors, and that the inclusion of 2022 introduces greater venue-level
variability, likely due to systematic differences in RTs that year.

\begin{figure}[tbp]
  \centering
  \includegraphics[width=\textwidth]{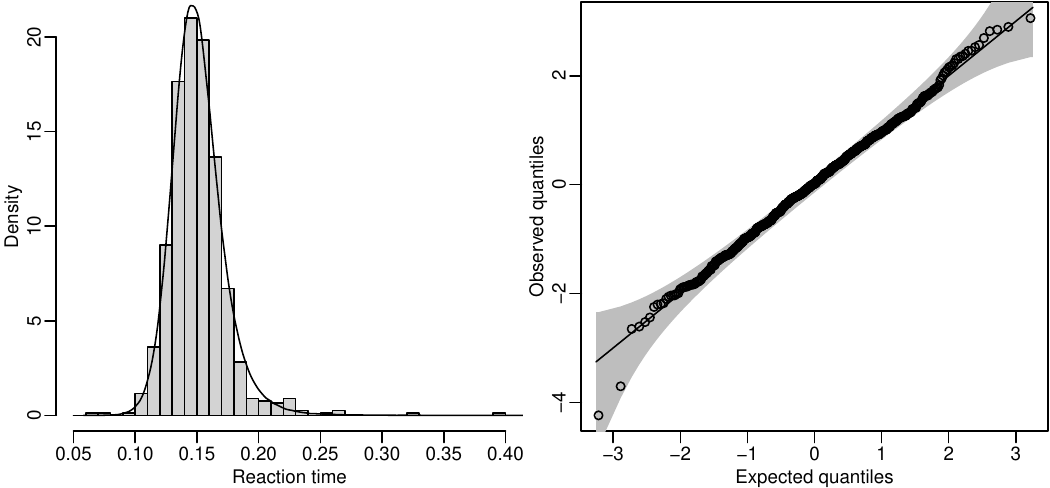}
  \caption{Diagnosis of the fitted GG distribution with 2022 included and
    random effects in model parameters: kernel density of 1 million
    observations drawn from the fitted model overlaid with the
    histogram of the observed RTs (left); Q-Q plot of the
    normal z-score of the quantile residuals from the fitted model (right).}
  \label{fig:diagnosis}
\end{figure}

Figure~\ref{fig:diagnosis} presents diagnostic checks for the fitted
GG distribution model with random effects. \eds{Is Fig 3 for the analysis with 
or without 2022?}
\of{With 2022. I added that to the caption of the figure} The left
panel compares the kernel density estimate of one million simulated RTs
from the fitted model to the histogram of the observed RTs.
The close alignment between the density curve and the histogram suggests that
the fitted GG model adequately captures the overall distribution of the
RTs. The right panel shows a Q-Q plot of the z-scores of the
quantile residuals from the fitted model. The points lie approximately along
the 45-degree reference line, indicating that the residuals are consistent
with the standard normal distribution, supporting the adequacy of the model
fit. These diagnostics collectively demonstrate that the fitted model provides
a reasonable representation of the observed RT data.

\begin{table}
  \centering
  \caption{Probabilities of observing RTs less than threshold 0.08,
  0.09, and 0.10 seconds based on the
    fitted GG GAMLSS model with both venue- and heat-level
random effects.}
  \begin{tabular}{c c c c}
   \toprule
   Data Set & Threshold 0.08 & Threshold 0.09 & Threshold 0.10  \\
   \midrule
   Excluding 2022 & $5.31\cdot10^{-5}$ & $3.53\cdot10^{-4}$ &  $1.94\cdot10^{-3}$  \\
   Including 2022 & $6.84\cdot10^{-5}$ & $4.95\cdot10^{-4}$ & $2.76\cdot10^{-3}$ \\
   \bottomrule
  \end{tabular}
  \label{tab:Sim_probability}
\end{table}

The fitted GG GAMLSS model with both venue- and heat-level
random effects provides a framework for assessing how extreme RTs
below certain thresholds are. The probability of observing a RT
below a given threshold, assuming no intentional false starts, was approximated
by generating 10 million realizations from the fitted model.
Table~\ref{tab:Sim_probability} summarizes the probabilities of observing
RTs below 0.08, 0.09, and 0.10 seconds under two scenarios: one
excluding and the other including data from 2022. Excluding 2022 slightly
reduces the probability of observing a fast RT, but the difference
is small. For example, the probability of a RT below 0.10 seconds
decreases from $2.76 \cdot 10^{-3}$ (approximately one in 362 starts) to
$1.94 \cdot 10^{-3}$ (approximately one in 515 starts) when 2022 is excluded.
Lowering the RT threshold from 0.10 to 0.08 seconds drastically
reduces the likelihood of observing a RT below the barrier, with
the probability dropping from one in every 362 starts (at 0.10 seconds) to one
in every 14620 starts (at 0.09 seconds) and one in every 146198 starts (at 0.08
seconds) when 2022 is included. These results highlight the rarity of extremely
fast RTs and substantiate the recommendations of \citet{komi2009iaaf}
to carefully consider the selection of RT thresholds.

Utilizing the same model, we can determine suitable RT barriers
based on the probability of observing a time below that barrier. As shown in
Table~\ref{tab:Sim_time}, including the 2022 data suggests a RT
barrier of 0.094 seconds to maintain a 0.1\% chance of observing an
exceptionally fast RT, while a stricter threshold of 0.082 seconds
is needed to limit this probability to 0.01\%. Excluding the 2022 data results
in slightly higher thresholds of 0.096 and 0.083 seconds for the respective
probability levels. These results indicate that while the inclusion of 2022
data slightly reduces the recommended barrier, the magnitude of the difference
is relatively small. This approach allows for tailoring RT
thresholds to desired levels of false positive rates, balancing fairness and
precision in disqualification criteria.

\begin{table}
  \centering
  \caption{Suggested RT barriers based on tail probabilities.}
  \begin{tabular}{c c c c}
   \toprule
   Data Set & Tail probability  $10^{-2}$ & Tail probability  $10^{-3}$ & Tail probability $10^{-4}$ \\
   \midrule
   Excluding 2022 & $0.111$ & $0.096$ & $0.083$ \\
   Including 2022 & $0.108$ & $0.094$ & $0.082$ \\
   \bottomrule
  \end{tabular}
  \label{tab:Sim_time}
\end{table}

\section{Discussion}\label{sec:disc}

This study first examined whether reaction times (RTs) at the 2022
World Track and Field Championships were significantly faster than at
other competitions. Our analyses indicate that RTs at the 2022 World
Championships were consistently faster than those recorded at both
national-level competitions earlier in the same year and the 2019 and
2023 World Championships. The persistence of this pattern across
different comparison groups suggests that these differences are not
due to random variation or individual improvements over time. A more
comprehensive analysis would benefit from a centralized database
containing RTs from all World Athletics-certified meets, but such data
are not consistently available. However, by incorporating competitions
from multiple years (2019, 2022, and 2023), the analysis accounts for
potential confounding factors such as seasonality and age, as athletes
at different stages of their careers are represented in different
comparisons.

This study further assessed whether the 0.1-second RT threshold is a
fair standard for disqualification. Our analyses of the GAMLSS model
suggest that while RTs below 0.1 seconds are rare, they may not be
as extraordinary as traditionally assumed. For men,
Table~\ref{tab:Sim_time} shows that RTs below 0.1 seconds
occur with a probability of approximately one in 362 starts when
including the 2022 data. Lowering the threshold to 0.08 seconds
drastically reduces this likelihood, supporting the idea that the
current barrier could be adjusted to reflect more realistic
probabilities of false starts. A similar pattern is observed for
women, as detailed in the Supplementary Material, where RTs
 below 0.1 seconds are exceedingly rare for the 100-meter dash
and 100-meter  hurdles. However, the uniformity of the 0.1-second
barrier for both men and women  is questionable, given numerous
studies documenting gender differences in RTs
\citep[e.g.,][]{lipps2011implications, babicc2009reaction,
  panoutsakopoulos2020gender}. These studies suggest that the current
threshold may  unfairly penalize men, for whom sub-0.1-second RTs are more 
probable. \citet{brosnan2017effects} advocate for
gender-specific barriers, a position that  aligns with our findings
and highlights the importance of tailoring thresholds to biological
distinctions.

This study provides a statistical framework to examine Devon Allen’s
disqualification at the 2022 World Track and Field Championships,
offering insights rather than drawing definitive conclusions about
potential equipment malfunction. Our findings indicate that RTs
 at the 2022 World Championships were, on average, faster than at other
competitions, as evidenced by the significant p-values in
Table~\ref{tab:Clusrankresults}. Additionally, the GAMLSS results
suggest that the 0.1-second barrier may not be as stringent as
previously believed. Based on Table~\ref{tab:Sim_time}, a stricter
threshold of 0.08 seconds could be considered, allowing athletes like
Allen to react swiftly without undue risk of disqualification. While
this analysis provides a rigorous statistical perspective, it does not
consider biomechanical factors, such as individual variability in
neuromuscular response times or the role of starting block sensors in
detecting pressure changes, which may offer more direct evidence of
reaction capabilities. In summary, while the results designate 2022 as
an anomalous year, Allen’s time, despite resulting in
disqualification, may not be categorically extreme.

\section*{Supplementary Material}
Additional results are summarized in a supplement for (1) rank-based
comparison with pooled (men and women) data, (2) GAMLSS results
for women's data, and (3) sensitivity of including positive yet disqualified
reaction times in GAMLSS.
The data and R code used for the analysis are available in a compressed file for
ease of reproducibility.

\bibliographystyle{apalike}
\bibliography{citations}

\end{document}


\title{\bf Supplement to ``On Devon Allen's Disqualification at the 2022 World Track and Field Championships''}

\if0\blind
{
  \author{Owen Fiore, 
  Elizabeth D. Schifano, 
  Jun Yan\\[1ex]
  Department of Statistics, University of Connecticut\\
}
} \fi

\if1\blind
{
  \bigskip
  \bigskip
  \bigskip
  \author{Anonymous Authors}
  \bigskip
} \fi

\maketitle 

\doublespace

\section{Rank-Based Comparison with Pooled Data}

As an alternative to the methods described in Section 3.1 of the main paper, it
is possible to combine all men's and women's reaction time (RT) data for each of 
the three competition
comparisons.  Thus we shrink our analyses from six to three, but each analysis
is roughly twice as big as previously.
For the 2022 national versus
international comparison, there were 160 RTs from 35
athletes and the asymptotic test result was a p-value of $1.94 \cdot 10^{-7}$.
For the 2019 versus 2022 comparison,
there were 258 RTs from 65 athletes and the asymptotic
test result was a p-value of $3.56 \cdot 10^{-8}$. 
For the 2023 versus 2022 comparison, there were 343 RTs
from 92 athletes and the asymptotic test result was a p-value of
$4.99 \cdot 10^{-12}$.  As a result of the larger sample sizes, the 
permutation-based tests were very computationally expensive to run and so we 
used a smaller number of permutations ($100,000$ instead of $1,000,000$).  All
three permutation tests had a p-value of $1 \cdot 10^{-5}$, which is the smallest
possible value given $100,000$ permutations.  When taken together
with the asymptotic results, the message is very clear. These are all
highly significant test results that show substantial differences in average RT
for athletes competing at multiple championship-level competitions.

\section{GAMLSS Results for Women Data}

\begin{figure}[tbp]
  \centering
  \includegraphics[width=\textwidth]{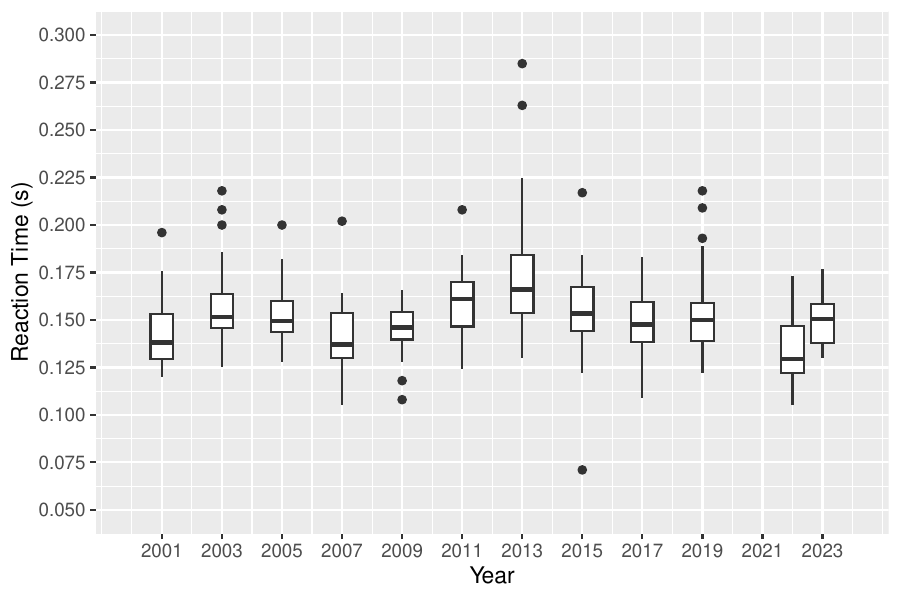}
  \caption{The RTs from 2021 to 2023 for the women's 100 meter hurdle
  and 100 meter dash.}
  \label{fig:WomensBoxplot}
\end{figure}

We also apply the RT barrier analysis described in Section 3.2 to women's data,
fitting the same model to women's RT data from 2001 to 2023.
The RT data for women is visualized in Figure~\ref{fig:WomensBoxplot}.
Similar to the men's data, RTs from 2022 appear lower than in other
years. After removing one obvious outlier, which was a disqualified
reaction time, the same model for men's data fits the women's data
reasonably well.

\begin{table}
  \centering
  \caption{Estimated fixed-effect parameters with standard errors in
    parentheses and estimated standard deviations of the random effects from the men's and
    women's fitted GG distribution with venue level random
    effects in $\mu$ and heat level random effects in $\sigma$. $n$ denotes
    size of the data.}
  \label{tab:womensfit}
  \begin{tabular}{c c c c c c c}
    \toprule
    Dataset & $n$ & $\beta_0$ & $\gamma_0$ & $\nu$ & $\tau_v$ & $\tau_h$ \\
    \midrule
    Women's & 732 & $-$1.921 (0.007) & $-$2.071 (0.028) & $-$3.691 (0.665) & 0.057 & 0.111 \\
    Men's & 776 & $-$1.910 (0.005) & $-$2.200 (0.027) & $-$1.178 (0.447) & 0.058 & 0.320 \\
    \bottomrule
  \end{tabular}
\end{table}

The fitted parameters in comparison with those from men's data are
summarized in Table~\ref{tab:womensfit}.  One notable aspect is that while
the venue effect standard deviation is nearly identical, the women's heat effect
standard deviation is smaller and $\nu$ is much larger.
This suggests that men’s races exhibit greater 
variability in RTs, possibly due to faster athletes influencing others 
to react more quickly in certain instances. The consistency in the venue effect 
standard deviation across men’s and women’s data indicates that the venue effect 
is not only statistically significant but also consistent in magnitude across 
genders. These findings highlight the potential impact of competition dynamics 
on heat variability and the robustness of venue-level effects.

\begin{table}
  \centering
  \caption{Probabilities of observing RTs less than threshold 0.08,
  0.09, and 0.10 seconds based on the men's and women's
    fitted GG GAMLSS model with both venue- and heat-level
random effects.}
  \begin{tabular}{c c c c}
   \toprule
   Data Set & Threshold 0.08 & Threshold 0.09 & Threshold 0.10  \\
   \midrule
   Women's & $1\cdot10^{-7}$ & $1.12\cdot10^{-5}$ &  $5.46\cdot10^{-4}$  \\
   Men's   & $6.84\cdot10^{-5}$ & $4.95\cdot10^{-4}$ & $2.76\cdot10^{-3}$ \\
   \bottomrule
  \end{tabular}
  \label{tab:Sim_prob_women}
\end{table}

We also repeat the simulation methods described in the paper to evaluate
the probability of an extreme RT for women.  Table~\ref{tab:Sim_prob_women}
compares the men's and women's results of observing RTs less than
0.08, 0.09, and 0.1 seconds.  We find across all three thresholds that
women have a lower probability of having a fast RT. The results are slightly
different from those from the men's data, which echoes existing studies
reporting gender differences in RTs \citep[e.g.,][]{lipps2011implications,
babicc2009reaction, panoutsakopoulos2020gender}.

\begin{table}
  \centering
  \caption{Suggested RT barriers based on tail probabilities.}
  \begin{tabular}{c c c c}
   \toprule
   Data Set & Tail probability  $10^{-2}$ & Tail probability  $10^{-3}$ & Tail probability $10^{-4}$ \\
   \midrule
   Women's & $0.111$ & $0.102$ & $0.095$ \\
   Men's   & $0.108$ & $0.094$ & $0.082$ \\
   \bottomrule
  \end{tabular}
  \label{tab:Sim_time_women}
\end{table}

To evaluate a fair RT barrier for women, we compared suggested 
barriers for men and women based on tail probabilities, as shown in 
Table~\ref{tab:Sim_time_women}. The results indicate that men are more likely 
to be disqualified under the current uniform 0.1-second threshold due to their 
generally faster RTs. This suggests that the same RT standard may not 
have equivalent implications for men and women. Potential adjustments could 
involve raising the barrier for women to align with men’s disqualification rates 
or lowering the barrier for men to match women’s rates. However, further research 
is needed to validate these findings and explore their broader implications. These 
results contribute to ongoing discussions about RT thresholds and emphasize the 
importance of statistical evidence in guiding decisions about competition fairness.

\section{Results from Data Excluding Positive Disqualified RTs}

An earlier iteration of the paper fit a model that did not include RTs from
athletes who were disqualified or did not finish but still registered a RT.  
However, it was ultimately decided to include these times to better estimate the 
left tail of the distribution and more accurately predict the probability of a 
low RT, as described in the main paper. 
We did not include negative RTs, however, as these represent a mistake of the
runner for starting before the gun is fired and are thus meaningless in our
objective to determine a fair RT barrier.  Not all of those disqualified were
disqualified because of breaking the 0.1 reaction time barrier; there are many
reasons why an athlete may be disqualified, with the most notable being failed
drug tests and lane violations.  

Nonetheless, in this section,
we exclude all disqualified RTs to see their effect on the probability of an
extreme RT. We otherwise fit an identical generalized Gamma model to the men's 
dash and hurdles RT data (including 2022), as presented in Section 3.2,
to determine how sensitive our model is to the inclusion/exclusion of these times.

\begin{table}
  \centering
   \caption{Probabilities of observing RTs less than threshold 0.08,
   0.09, and 0.10 seconds based on the
     fitted GG GAMLSS model with both venue- and heat-level
 random effects.}
   \begin{tabular}{c c c c}
    \toprule
    Data Set & Threshold 0.08 & Threshold 0.09 & Threshold 0.10  \\
    \midrule
    Without DQs & $4.93\cdot10^{-5}$ & $3.53\cdot10^{-4}$ &  $1.97\cdot10^{-3}$  \\
    With DQs & $6.84\cdot10^{-5}$ & $4.95\cdot10^{-4}$ & $2.76\cdot10^{-3}$ \\
    \bottomrule
   \end{tabular}
   \label{tab:DQSim_probability}
 \end{table}

Table~\ref{tab:DQSim_probability} shows the effect of removing disqualified (DQ)
times from the analysis.  The probability of observing extreme RTs is lower when
we remove the 17 observations.  While the probability of observing an extreme
RT is less when we remove the RTs of disqualified athletes, many of the 
conclusions remain the same: the current standards for disqualification are not
grounded in statistical analysis, and there appear to be unequal standards for men
and women.

\bibliographystyle{apalike}
\bibliography{citations}